\documentclass[chicago]{emulateapj}
\usepackage{amsmath}
\usepackage{graphics,graphicx}

\shorttitle{Subaru-HiCIAO survey of FU Orionis objects}
\shortauthors{Liu et al.}
\slugcomment{Accepted to Science Advances, Nov.22.2015}
\begin{document}

\title{Circumstellar Disks of the Most Vigorously Accreting Young Stars \footnote{Based [in part] on data collected at Subaru Telescope, which is operated by the National Astronomical Observatory of Japan}}

\author{Hauyu Baobab Liu\altaffilmark{1,2}} \author{Michihiro Takami\altaffilmark{1}} \author{Tomoyuki Kudo\altaffilmark{3}} \author{Jun Hashimoto\altaffilmark{3}} \author{Ruobing Dong\altaffilmark{1,4,5}} \author{Eduard I. Vorobyov\altaffilmark{6,7}} \author{Tae-Soo Pyo\altaffilmark{8}} \author{Misato Fukagawa\altaffilmark{3}} \author{Motohide Tamura\altaffilmark{9}} \author{Thomas Henning\altaffilmark{10}} \author{Michael M. Dunham\altaffilmark{11}} \author{Jennifer Karr\altaffilmark{1}} \author{Nobuhiko Kusakabe\altaffilmark{3}} \author{Toru Tsuribe\altaffilmark{12}}

\affil{$^{1}$Academia Sinica Institute of Astronomy and Astrophysics, P.O. Box 23-141, Taipei, 106 Taiwan (baobabyoo at gmail. com)} 
\affil{$^{2}$European Southern Observatory (ESO), Karl-Schwarzschild-Str. 2, D-85748 Garching, Germany} 
\affil{$^{3}$National Astronomical Observatory of Japan, 2-21-1 Osawa, Mitaka, Tokyo 181-8588 Japan}
\affil{$^{4}$Hubble Fellow}
\affil{$^{5}$Department of Astronomy, UC Berkeley, 147 Del Mar Ave, Berkeley, CA, 94708}
\affil{$^{6}$Department of Astrophysics, University of Vienna, Tuerkenschanzstrasse 17, 1180, Vienna, Austria}
\affil{$^{7}$Research Institute of Physics, Southern Federal University, Rostov-on-Don, 344090, Russia}
\affil{$^{8}$Subaru Telescope, National Astronomical Observatory of Japan, 650 North Aohoku Place Hilo, HI 96720, USA}
\affil{$^{9}$Department of Astronomy, Graduate School of Science, The University of Tokyo, 7-3-1, Hongo, Bunkyo-ku, Tokyo 113-0033, Japan}
\affil{$^{10}$Max-Planck-Institut f\"{u}r  Astronomie K\"{o}nigstuhl, 17 D-69117 Heidelberg}
\affil{$^{11}$Harvard-Smithsonian Center for Astrophysics, 60 Garden St, MS 78, Cambridge, MA 02138}
\affil{$^{12}$College of Science, Ibaraki University, Bunkyo 2-1-1, Mito, 310-8512 Ibaraki, Japan}

\def\Ms{$M_{\rm *}$}
\def\Lsun{$L_{\odot}$}
\def\Hi{H\,{\sc i}~}				
\def\Ht{H$_2$\,}                    
\def\COto{$^{12}$CO(J=2$\rightarrow$1)\ }
\def\cch{C$_2$H }
\def\c2h2{C$_2$H$_2$}
\def\hcccn{HC$_3$N~}

\def\kms{$~\rm km\,s^{-1}$}
\def\micro{\,$\mu$m}
\def\Msun{$M_{\odot}$}

\begin{abstract}
Young stellar objects (YSOs) may not accumulate their mass steadily, as was previously thought, but in a series of violent events manifesting themselves as sharp stellar brightening. 
These events can be caused by fragmentation due to gravitational instabilities in massive gaseous disks surrounding young stars, followed by migration of dense gaseous clumps onto the star. 
We report our high angular resolution, coronagraphic near-infrared polarization imaging observations using the High Contrast Instrument for the Subaru Next Generation Adaptive Optics  (HiCIAO) of the Subaru 8.2 m Telescope, towards four YSOs which are undergoing luminous accretion outbursts.
The obtained infrared images have verified the presence of several hundred AUs scale arms and arcs surrounding these YSOs. 
In addition, our hydrodynamics simulations and radiative transfer models further demonstrate that these observed structures can indeed be explained by strong gravitational instabilities occurring at the beginning of the disk formation phase. 
The effect of those tempestuous episodes of disk evolution on star and planet formation remains to be understood.
\end{abstract}



\section{Introduction} \label{sec:introduction}
The formation of solar-like systems and binary stars may not simply follow the quasi-stationary paradigm of classical analytical calculations (Shu 1997; Young \& Evans 2005). 
Instead, protoplanetary disks around these objects might experience an extremely chaotic evolutionary process, whereby vigorous protostellar mass-accumulation and disk stabilization are via inward migration and ejection of large spiral arcs and massive fragments. 
In fact, the observed luminosity of protostars is far less than the expected luminosity inferred from their averaged accretion rate (Evans et al. 2009). 
The most promising solution to this apparent paradox is episodic accretion (Kenyon et al. 1993a,b), which is supported by the observation of extreme variations (4-6 magnitudes) in the optical/infrared brightness of some YSOs (Herbig 1989; Hartmann \& Kenyon 1996). 
However, these so-called \textit{accretion outburst} sources are not yet fully understood. 

\begin{figure*}
\hspace{0.25cm}
\begin{tabular}{ p{17cm}  }
\includegraphics[width=17cm]{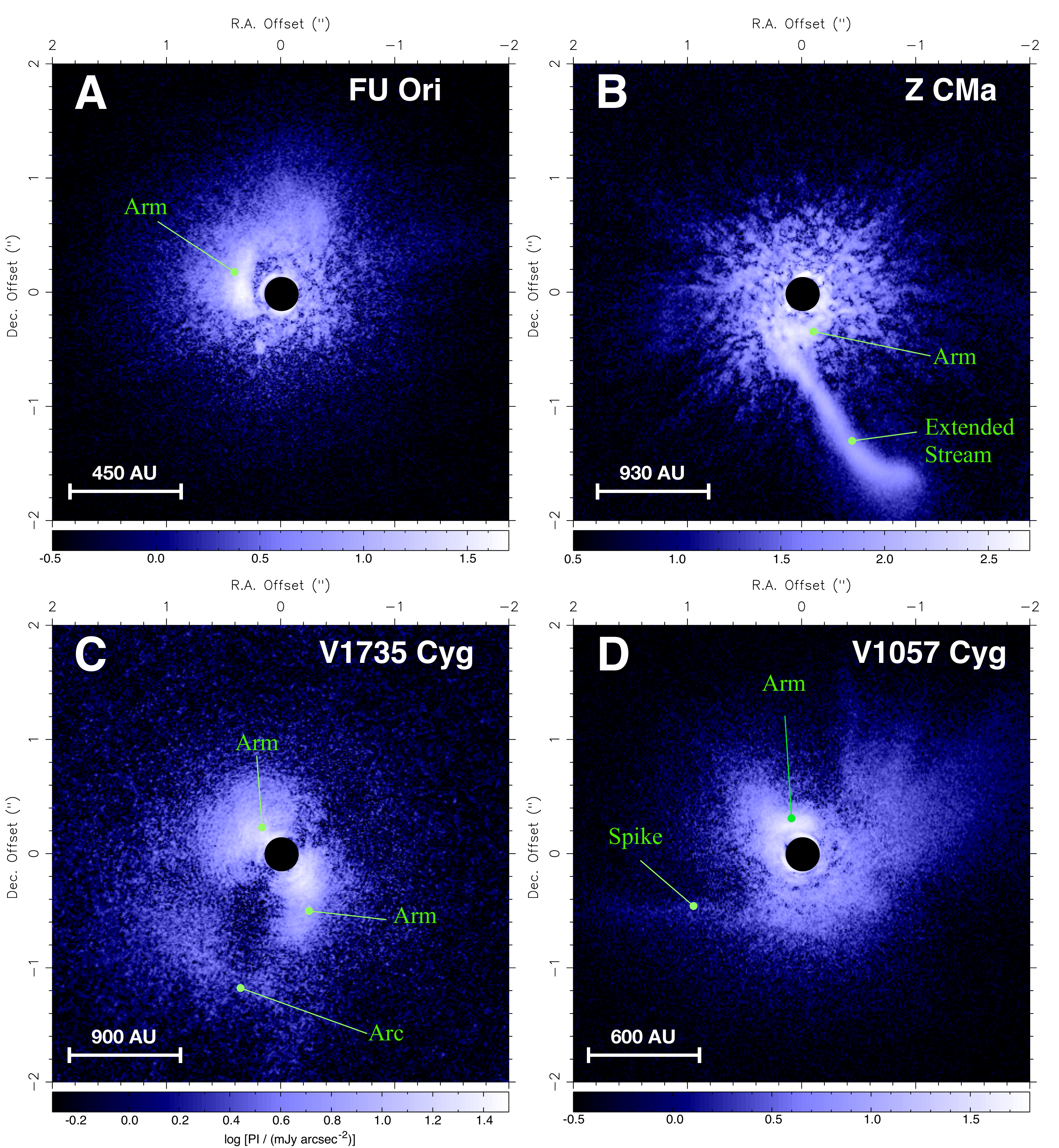}  \\
\end{tabular}
\caption{\footnotesize{The polarized intensity images of the selected FU Orionis objects. 
Panels A, B, C, D present the Subaru-HiCIAO H-band image of FU Ori, the K-band image of Z CMa, the K-band image of V1735 Cyg, and the H-band image of V1057 Cyg, respectively. 
Scale bars in panels A-D are based on the assumption of distances of 450 pc, 930 pc, 900 pc, and 600 pc. 
The brightness is presented using a log scale. 
The central 0$''$.3 scale region is masked in each panel (see Section \ref{subsec:observation}). 
Annotations of structures are given for convenience; their relations to the distribution of gas are discussed in the main text.
}} 
\vspace{0.8cm}
\label{fig:hiciao}
\end{figure*}

FU Orionis objects, also known as FUors, undergo accretion outbursts during which the mass accretion rate onto the star rapidly increases by a factor of $\sim$1000, and remains there for several decades or more. 
A sudden increase in the accretion rate heats up the inner disk (r$<$1AU), observable as enhanced continuum emission at optical and infrared wavelengths. 
The recent numerical hydrodynamics simulations of collapsing, $\sim$0.1 pc scale, $\sim$1 $M_{\odot}$ cores (Vorobyov \& Basu 2010; Machida et al. 2011; Dunham \& Vorobyov 2012; Vorobyov \& Basu 2015) have proposed that, in a short-lived phase after disk formation, mass infall from the collapsing cloud onto the disk usually exceeds mass accretion from the disk onto the protostar. 
The resulting increase in the mass of the disk destabilizes the disk, leading to the formation of spiral arms and fragments. 
These gravitationally unstable disks can exhibit FU-Orionis-type outbursts when fragments are driven onto the protostar via gravitational interactions with other fragments or spiral arms. 
In addition, multiple massive ($\le$0.1 $M_{\odot}$) forming fragments may settle in quasi-stable wide-separation orbits or be torqued into the disk inner regions, which naturally explains the detections of massive planets at a broad range of separations from the host star (Nayakshin 2010; Vorobyov \& Basu 2010; Machida et al. 2011; Kuzuhara et al. 2013; Vorobyov 2013). 
Moreover, the violent protostellar accretion in embedded phase via gravitational instability may explain the absence of an enhanced ionized magnetohydrodynamic (MHD) jet from the recently report Class 0 FU Orionis candidate, HOP 383 (Galv\'{a}n-Madrid et al. 2015).
Gravitational perturbations from close encounters with (sub)stellar companions or chance encounters with dense stellar clusters may add to the complexity of structures and can trigger accretion outbursts (Bonnell \& Bastien 1992; Pfalzner 2008), although these mechanisms can only operate in non-isolated systems.
Other outburst triggering mechanisms include the planet-disk interaction (Nayakshin \& Lodato 2012) and the thermal instability in the inner disk (Lin, Papaloizou \& Faulkner 1985; Bell \& Lin 1994). However, the recent discovery of FU-Orionis-type outbursts from Class 0/I YSOs (Caratti o Garatti et al. 2011; Safron et al. 2015) seems to require planet formation much earlier than is presently accepted. The thermal instability has difficulty with explaining the observational facts (Zhu et al. 2007; Zhu et al. 2009a) and is now superseded by a more elaborate model combining gravitational instability in the outer disk plus the magnetorotational instability in the inner disk (Zhu et al. 2009b; Bae et al. 2014).

\begin{figure}
\hspace{-0.25cm}
\begin{tabular}{ p{10cm}  }
\includegraphics[width=8.7cm]{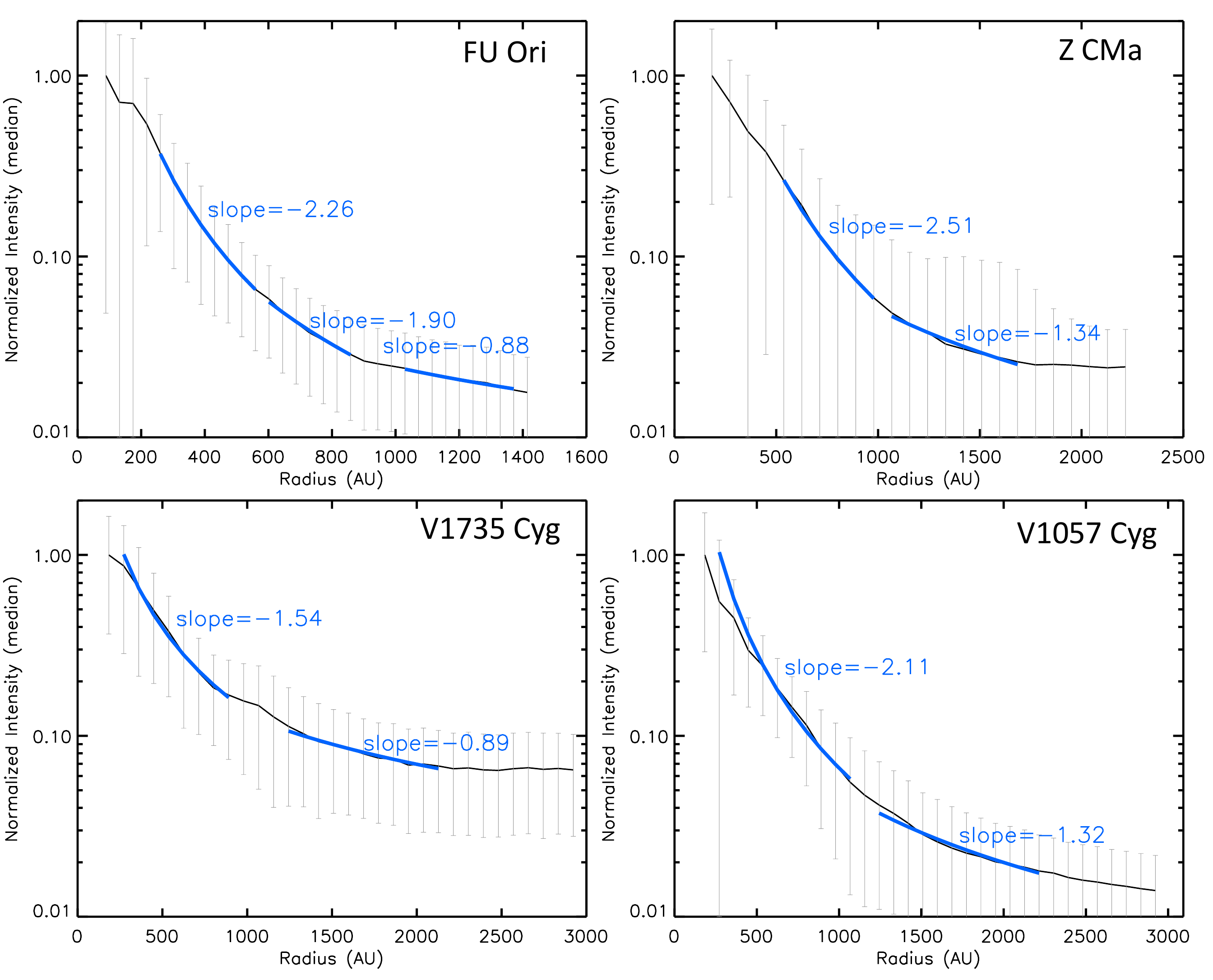}  \\
\end{tabular}
\vspace{-0.0cm}
\caption{\footnotesize{
Normalized radial intensity profiles (black) of images presented in Figure \ref{fig:hiciao}. 
The intensities shown in each panel are medians derived from 10 pixels wide annuli, which were normalized so to a maximum of 1.0. 
We fitted sectors of these profiles assuming the intensity is proportional to $r^{slope}$, where $r$ is the radius from the host stars. 
Blue line segments show the results of the fits.
}} 
\vspace{0.8cm}
\label{fig:profile}
\end{figure}

To gain insight into this scenario, we have performed Subaru-HiCIAO (Tamura et al. 2006) H-band and K-band (i.e. 1.6 and 2.2 $\mu$m) polarization differential imaging (PDI) towards 4 of the 11 confirmed FUors with optical and infrared outbursts (we refer to a review article Audard et al. 2014 for a summary of these sources): FU Ori, V1735 Cyg (also known as Elias 1-12), V1057 Cyg, and Z CMa. 
The obtained linear polarization intensity images (PI) are sensitive to stellar light reflected by tiny amounts of dust (e.g. the order of the moon mass, e.g. Takami et al. 2013). 
Therefore, these observations provide a powerful tool for understanding surface morphology of circumstellar disks, picking out relics of interactions without being seriously confused by dominant gas structures, and for identifying residual envelopes surrounding the YSOs (e.g. Takami et al. 2014). 
In addition, we have performed numerical hydrodynamics simulations and radiative transfer modeling, to compare with the Subaru images. 
Our observations, numerical hydrodynamics simulations, and radiative transfer modeling are outlined in Section \ref{sec:observation}.
The results are presented in Section \ref{sec:result}.
The physical implications are discussed in Section \ref{sec:discussion}.
Our conclusion is give in Section \ref{sec:summary}. 


\section{Observations and Modeling} \label{sec:observation}
\subsection{Subaru-HiCIAO Observations} \label{subsec:observation}
The linear polarization differential imaging (PDI) technique was used for obtaining the polarized intensity (PI) images. 
We carried out observations towards Z CMa at  J-band (1.3 $\mu$m), FU Ori, V1735 Cyg, V1057 Cyg and Z CMa at H-band (1.6 $\mu$m), and V1735 Cyg and Z CMa at K-band (2.2 $\mu$m) on 2014 October 04 , 05, and 06, using Subaru-HiCIAO. 
The innermost working angle of these coronagraphic images is 0$''$.3. These observations were performed using the adaptive optics system (AO188; Hayano et al. 2004), and were combined with dual-beam polarimetry. For the H band images of FU Ori and V1057 Cyg, and K band images of V1735 Cyg and Z CMa used in the present letter, the achieved resolutions (FWHM) are $\sim$80, $\sim$80, $\sim$90, and $\sim$70 milli-arcsecond (mas), respectively. Our pixel scale is 9.5 mas. 
Of all observed images, the images selected to be presented here were taken under the best seeing and cloud conditions. 
Details of the full data product and more analysis will be described in a separate paper. 

We measured polarizations following the standard strategy of rotating the half-waveplate to four angular positions (in order 0$^{\circ}$, 45$^{\circ}$, 22$^{\circ}$.5, 67$^{\circ}$.5). 
Data reduction including de-striping, flat fielding, bad pixel removal, distortion correction, image re-alignment and the calculation of Stokes Q, U, and PI images followed a standard procedure outlined in our prior publications (Hashimoto et al. 2011). 
Our initial examination of the polarization vector images found that the bright polarized stellar halo, as a result of imperfect performance of adaptive optics, can contaminate the PI images. 
The primary effects are (i) the obvious alignment of polarization vectors along a specific position angle, and (ii) the cancellation of polarized reflection light from disk, in the position angles that the polarization vectors align. 
Therefore, we used a polarized halo-subtraction algorithm to suppress this contamination (Hashimoto et al. 2012). 
For all images, we estimated the polarization percentage of the stellar halo in annuli with inner and outer radii of 65 and 195 pixels. 
We removed a sky foreground/background averaged from annuli with inner and outer radii of 200 and 220 pixels. 
A comparison of the PI images before and after polarized halo-subtraction is provided in Appendix \ref{section:appendix}.
The other PI images presented in this paper have been halo-subtracted unless specifically indicated. 
We note that perfect modeling of the polarized stellar halo is challenging due to confusion with real polarized emission features. 
However, the subtle differences between the images before and after halo-subtraction are not critical to our discussion.

\begin{figure*}
\hspace{0.7cm}
\begin{tabular}{ p{16cm}  }
\includegraphics[width=16cm]{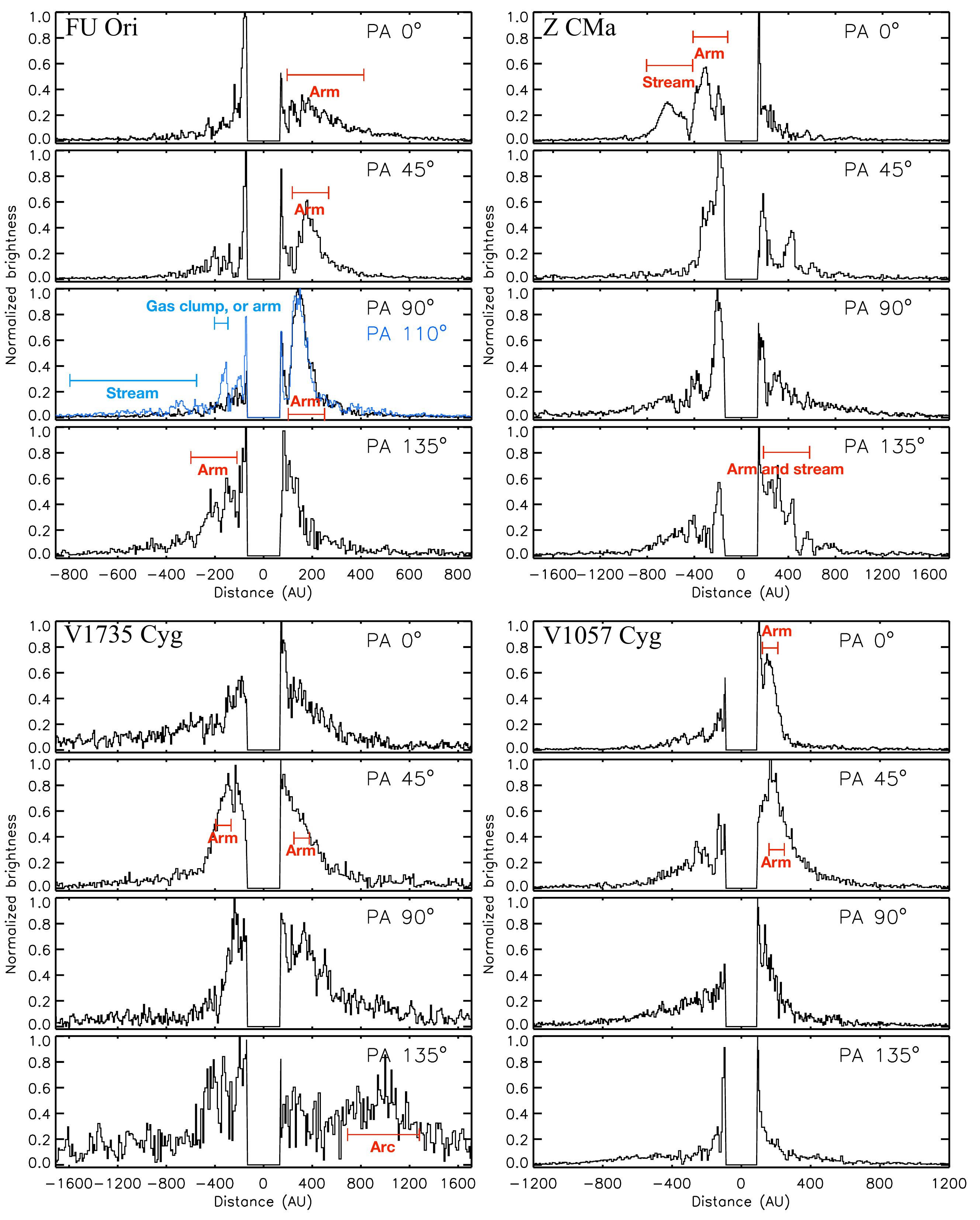}  \\
\end{tabular}
\caption{\footnotesize{Slices of the images presented in Figure 1, at the position angles (measured North through East) of 0$^{\circ}$, 45$^{\circ}$, 90$^{\circ}$, and 135$^{\circ}$. An additional slice at PA=110$^{\circ}$ is presented for FU Ori to demonstrate the faint and extended elongated structure in the northwest (see also Figure \ref{fig:smoothed}). Annotations highlight features from the images.
}} 
\vspace{0.8cm}
\label{fig:slice}
\end{figure*}

\subsection{Hydrodynamics Simulations and Radiative Transfer Modeling} \label{subsec:modeling}
The surface density and mid-plane temperature of the disk and inner envelope were determined using the numerical hydrodynamics code described in our prior papers (Vorobyov \& Basu 2010; Vorobyov \& Basu 2015). 
The code follows the gravitational collapse of a prestellar cloud core into the disk formation and evolution phase, in the thin-disk limit, and taking the dynamical infall of the parent core onto the disk's outer parts into account. 

Once formed, the protostellar disk occupies the inner part of the numerical polar grid, while the in-falling parental core occupies the outer part. 
The use of logarithmic scaling in the radial direction enables us to achieve sub-AU numerical resolution at distances $\le$100 AU and to fulfill the Truelove criterion (Truelove et al. 1998), which is important for the proper modeling of fragmenting disks (Vorobyov 2013). 
To prevent time steps from becoming too small, we introduce a "sink cell" at 5-10 AU and impose a free inflow inner boundary condition so that the matter is allowed to flow out of the computational domain but is prevented from flowing in. 
The sink cell is dynamically inactive; it contributes only to the total gravitational potential and ensures the smooth behavior of the gravitational force down to the stellar surface. 
During the early stages of the core collapse, we monitor the gas surface density in the sink cell and when its value exceeds a critical value for the transition from isothermal to adiabatic evolution, we introduce a central point-mass object representing the forming star. 
In subsequent evolution, 90\% of the gas that crosses the inner boundary is assumed to land on the central object. 
The other 10\% of the accreted gas is assumed to be carried away with protostellar jets.

\begin{figure}
\hspace{-0.25cm}
\begin{tabular}{ p{10cm}  }
\includegraphics[width=8.5cm]{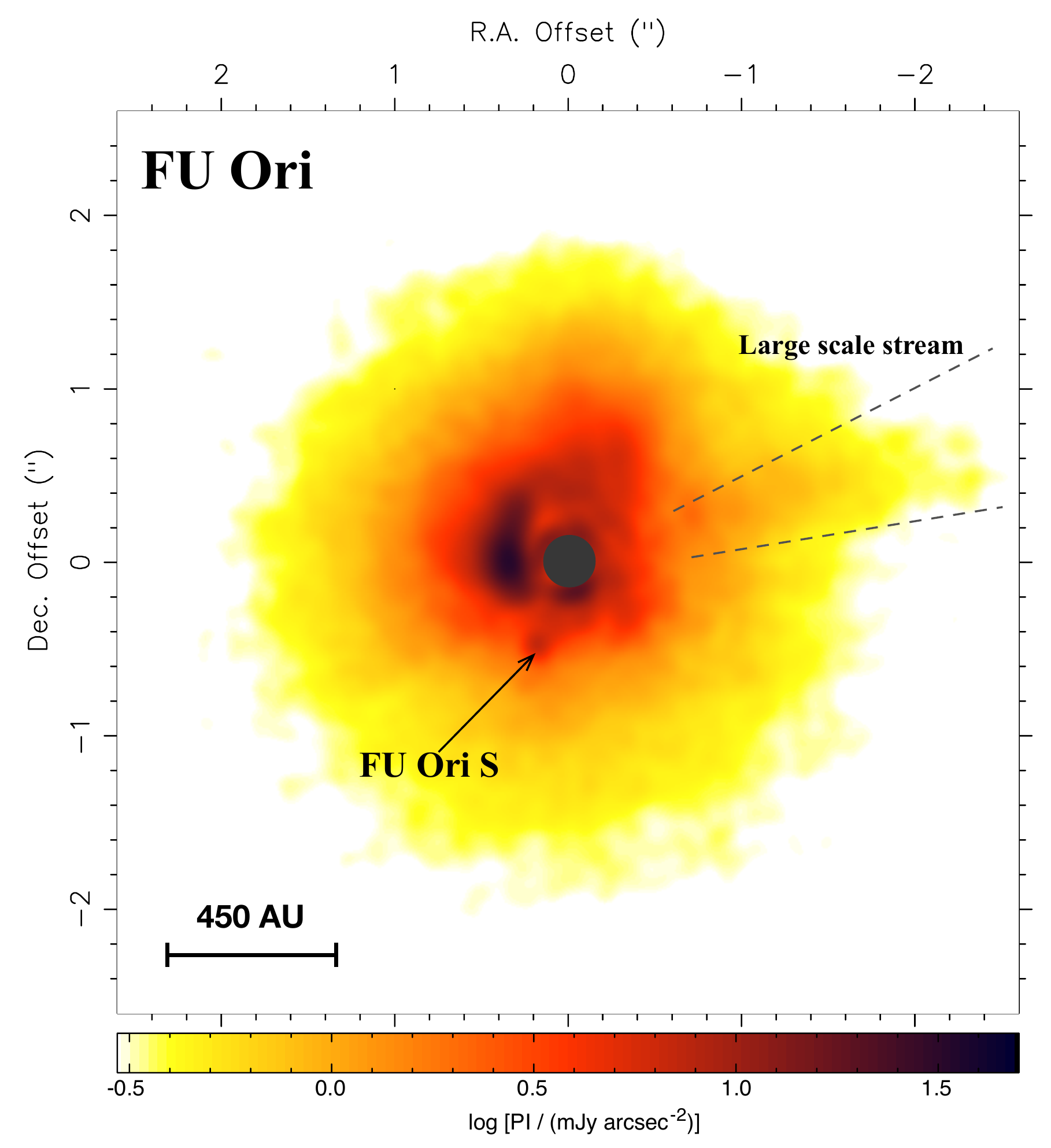}  \\
\end{tabular}
\vspace{-0.0cm}
\caption{\footnotesize{
The smoothed Subaru-HiCIAO H-band image of FU Ori, convolved with a two-dimensional Gaussian kernel with a standard deviation of 5 pixels. Color bar is log scaled. The location of the companion, FU Ori S is indicated by an arrow, although its brightness is diluted in this smoothed image (c.f. Figure \ref{fig:hiciao}).
}} 
\vspace{0.8cm}
\label{fig:smoothed}
\end{figure}

The code has been successfully applied to studying accretion rates, disk properties and gravitational fragmentation in circumstellar disks (Vorobyov 2010, 2011a,b; Dunham \& Vorobyov 2012; Basu \& Vorobyov 2012). 
It includes major physical and thermal processes involved in the formation of disks and planets, and solves for the full energy balance equation. 
Among the processes taken into account are disk cooling, stellar and background irradiation, viscous and shock heating, frozen-in magnetic fields, and disk self-gravity. 
After computing the integrated disk properties in the ($r$, $\phi$) plane, the vertical disk scale height is calculated using the assumption of hydrostatic equilibrium and taking into account both the gravity of the star and local disk gravity (Vorobyov \& Basu 2009). 
The two-dimensional disk structure plus the vertical scale height are later used to create the three-dimensional disk structure needed for the radiation transfer code (see below), using the local mid-plane temperature and assuming a Gaussian distribution in the vertical direction.

\begin{figure}[h]
\begin{tabular}{ p{10cm}  }
\includegraphics[width=8.5cm]{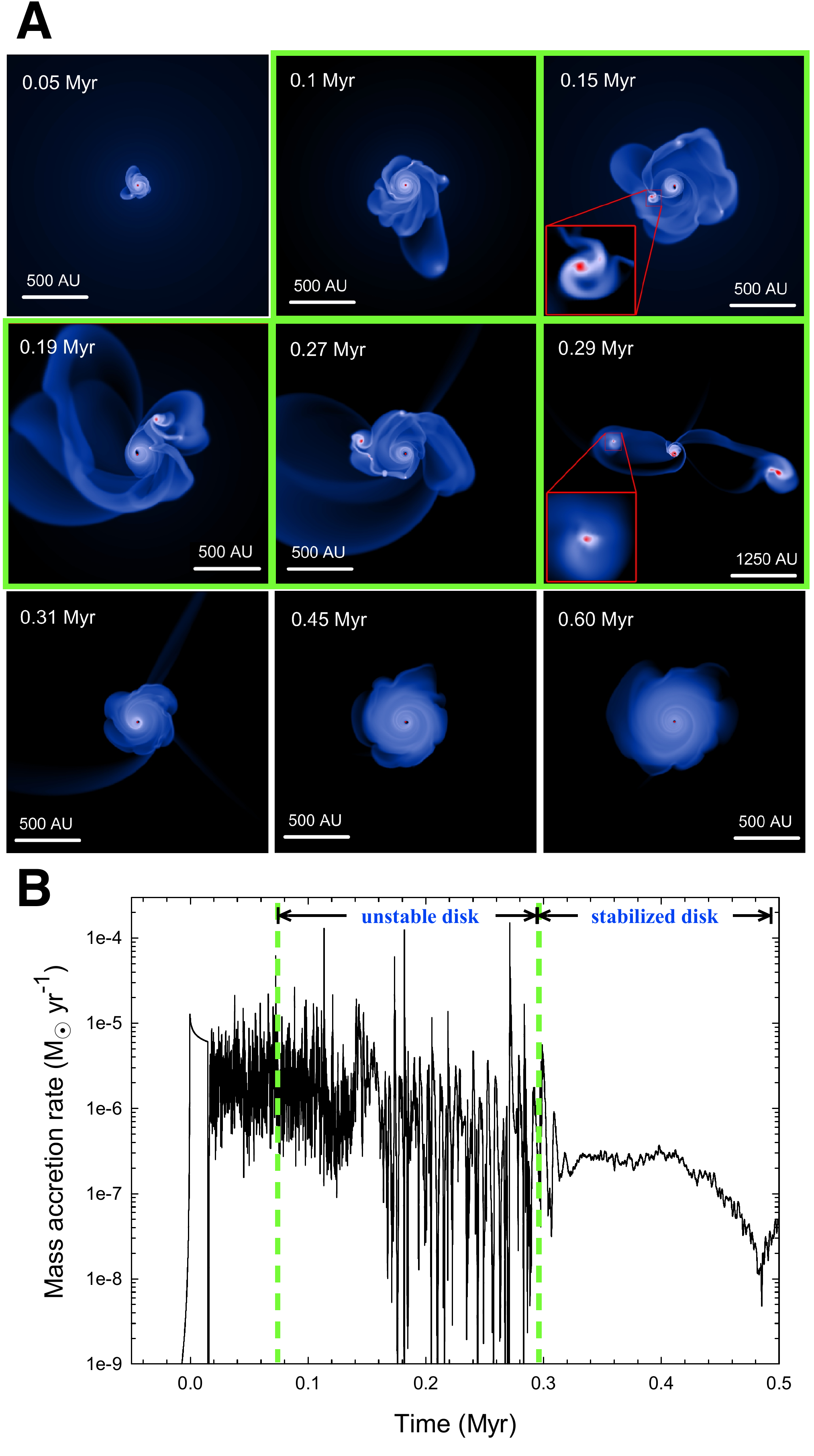}  \\
\end{tabular}
\vspace{-0.3cm}
\caption{\footnotesize{
(A) Simulated disk surface density, and the (B) protostellar accretion rate over the time period of the simulation. 
Green boxes in panel A, and the time period bound by the green dashed lines in panel B, represent the proposed evolutionary stage of our observed young stellar objects (Figure \ref{fig:hiciao}). 
The inset panels at 0.15 Myr highlight the most massive fragment formed in the simulation. The inset panel at 0.29 Myr indicates one of the two ejected fragments.
}} 
\vspace{0.8cm}
\label{fig:hydrodynamics}
\end{figure}

Our code was further improved (Vorobyov et al. 2013), by including the Lyon stellar evolution code to calculate the properties of the forming star (Chabrier \& Baraffe 1997), incorporated with the effects of accretion processes (Baraffe et al. 2009). 
The stellar evolution code is coupled with the main hydrodynamics code in real time. 
The input parameter to the stellar evolution code provided by the disk modeling is the mass accretion rate onto the star. 
The outputs of the stellar evolution code are the stellar radius and the photospheric luminosity, which are employed by the hydrodynamics disk simulations to calculate the total stellar luminosity and the flux of radiation reaching the disk surface.

We use the method of finite differences with a time-explicit solution procedure similar in methodology to the ZEUS code (Stones \& Norman 1992). 
The advection is treated using the third-order accurate piecewise-parabolic interpolation scheme (Colella \& Woodward 1984). 
The update of the internal energy due to cooling and heating is done implicitly using the Newton-Raphson method of root finding, complemented by the bisection method where the Newton-Raphson iterations fail to converge. The accuracy is guaranteed by not allowing the internal energy to change more than 30\% over one time step. 
A small amount of artificial viscosity is added to the code, though the associated artificial viscosity torques were shown to be negligible in comparison with gravitational torques (Vorobyov \& Basu 2007).

Models of reflected near-infrared light images, and the spectral energy distribution, are produced using the Whitney MCRT code (Whitney et al. 2013). 
This code has been used to model protoplanetary disks in the past (Dong et al. 2012; Hashimoto et al. 2012; Follette et al. 2013; Grady et al. 2013; Dong et al. 2014). 
In the MCRT simulations, the luminosity from the central star is scattered or absorbed and reemitted by the dust in the surrounding disk. 
We assumed the standard interstellar medium (ISM) dust grains, which contain silicate, graphite, and amorphous carbon (Kim et al. 1994). 
Their size distributions range from $\sim$0.02 to $\sim$1 $\mu$m, which can be approximated by a smooth power law distribution with $n(s)$$\sim$s$^{-3.5}$ in the range of 0.02-0.25 $\mu$m followed by a sharp cut off beyond 0.25 $\mu$m. 
Their optical properties can be found in our priori publication (Dong et al. 2012).
The temperature in each grid cell is calculated based on the radiative equilibrium algorithm (Lucy 1999). 
The anisotropic scattering phase function is approximated using the Henyey-Greenstein function. 
Polarization is calculated assuming a Rayleigh-like phase function for the linear polarization (White 1979). 
The 2D surface density and the mid-plane temperature profile determined by the hydrodynamics simulations are combined to create a 3D disk structure, assuming hydrostatic equilibrium and an isothermal temperature profile in the vertical direction. 
The 3D disk structure has 420$\times$200$\times$512 grid points in radial, polar and azimuthal directions, respectively, and the disk extends from the dust sublimation radius ($\sim$0.7 AU) to $\sim$500 AU. 
The MCRT simulations are run with 100 million photons to achieve a good S/N ratio. 
For simplicity, the central source is assumed to be a 0.6 $M_{\odot}$ protostar, with a temperature of 4300 K and a luminosity of 114 $L_{\odot}$, although in reality it can be a combination of sophisticated emission mechanisms including accretion shocks and the hot inner accretion flow/disk (Zhu et al. 2007, 2008). 
The density of the small (sub-$\mu$m-sized) grains responsible for reflected light is set to be proportional to the gas, as small grains are generally well mixed with the gas.
The opacity in the disk is determined by assuming a 100:1 gas-to-dust-mass ratio and a standard ISM dust opacity profile. 
The full resolution MCRT polarized intensity images produced by the Whitney code are convolved with a Gaussian point spread functions with a FWHM of 0$''$.07, to achieve a similar angular resolution to the Subaru observations. 
The simulated spectral energy distributions do not take into the consideration reddening effects owing to foreground extinction.


\section{Results} \label{sec:result}
Figure \ref{fig:hiciao} shows our Subaru-HiCIAO images. 
The radial intensity profiles of these images are presented in Figure \ref{fig:profile}.
We found that in the inner $\sim$1000 AU regions, the measured intensity profiles decrease with radius at rates ($r^{-(1.5\sim2.5)}$), which were generally observed in other disks at scales of a few ten to several hundred AUs (Kudo et al. 2008; Hashimoto et al. 2011; Akiyama et al. 2015). 
Radial profile of V1057 Cyg, however, is confused with spiky structures in the outer regions of its envelope. 
The radial intensity profiles drop less steeply on $>$1000 AU scales, which can be attributed to the presence of envelope material. 
The observed systems are likely protoplanetary disks embedded within complicated larger scale envelopes. 
The envelope of V1057 Cyg shows several spikes, connected with filaments on much more extended spatial scales. 
Those extended filaments were interpreted as a result of a relative velocity offset between the protostar and the ambient gas (Goodrich 1987). 

\begin{figure*}
\hspace{0.3cm}
\begin{tabular}{ p{15cm}  }
\includegraphics[width=16cm]{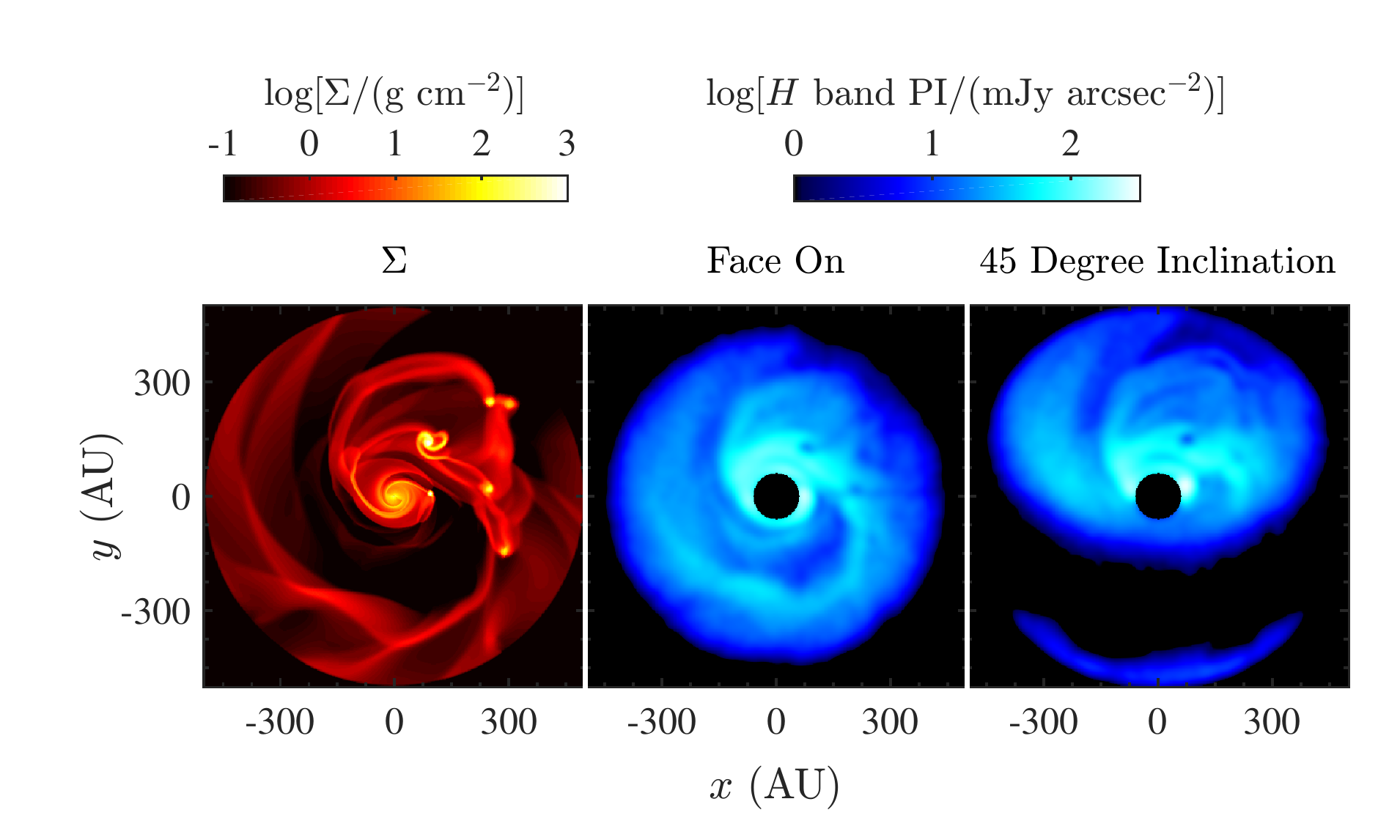}  \\
\end{tabular}
\vspace{-0.3cm}
\caption{\footnotesize{Radiative transfer models of reflected infrared light from a gravitationally unstable disk. 
Panels from left to right are the gas column density produced by hydrodynamics simulations, the simulated H-band reflected light image for a face on projection, and the simulated H-band reflected light image for the case of an inclination of 45$^{\circ}$. 
The image sizes are 1000 AU$\times$1000 AU. 
The simulated reflected light images have the same spatial resolution and occultation mask size as our Subaru-HiCIAO image of FU Ori.
}} 
\vspace{0.8cm}
\label{fig:toymodel}
\end{figure*}

We resolved a giant arm around FU Ori and its companion FU Ori S, stretching from east to northeast, at 50-500 AU scales. 
The other three sources are more distant, and therefore it is more difficult to resolve details. 
However, asymmetric disk features, which are consistent with giant arms, can be clearly seen in the south of Z CMa, the southwest and northeast of V1735 Cyg, and the north of V1057 Cyg. 
More than one arm, or gas clump, may appear to the west of FU Ori, which can be seen more clearly in the spatially smoothed PI image (Figure \ref{fig:smoothed}). 
The observed spiral arms show polarized intensity contrast of higher than $\sim$2-5 (Figure \ref{fig:slice}). 
The FU Ori and Z CMA disks are connected with $\sim$1000 AU scale, approximately radial elongated structures extending to the west and south, respectively. 
The bright linear feature in Z CMa was also reported in the previous observations (Millan-Gabet \& Monnier 2002; Canovas et al. 2015).
The PI image of V1735 Cyg has a large-scale arc, to the southeast. 

FU Ori and Z CMa have known close companions (Koresko et al. 1991; Wang et al. 2004). 
The companion of Z CMa is located within our 0$''$.3 occultation mask. 
The companion of FU Ori, FU Ori S, may be associated with a small circumstellar disk, thus can be seen in reflected light in our PI image (Figure \ref{fig:hiciao}). 
To our knowledge, there is no stellar companion within the fields of view shown here for V1735 Cyg or V1057 Cyg, although for all observed sources, unseen (sub-)stellar companions or massive gas clumps could be obscured by the optically thick disk. 
The morphologies resolved in our PI images, consisting of a combination of arms, extended envelope features, elongated structures and companions, are in excellent agreement with the unstable disk accretion scenario and the hydrodynamics simulations discussed earlier (Dunham \& Vorobyov 2012; Vorobyov \& Basu 2015), which can account for the intense accretion outbursts as well.

\section{Discussion} \label{sec:discussion}
\begin{figure}
\hspace{-0.2cm}
\begin{tabular}{ p{10cm}  }
\includegraphics[width=9.8cm]{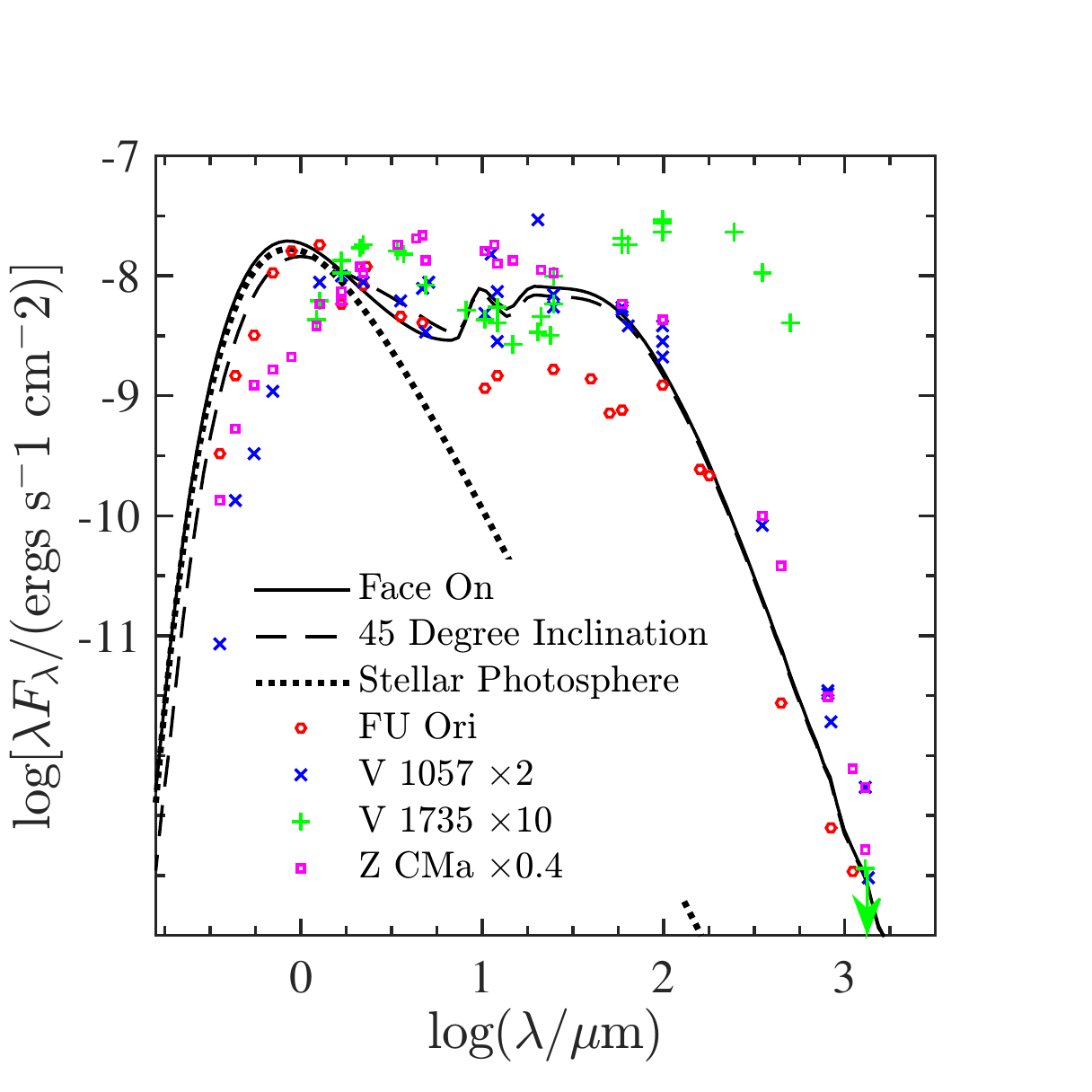}  \\
\end{tabular}
\vspace{-0.3cm}
\caption{\footnotesize{
The simulated spectral energy distributions for the model shown in Figure \ref{fig:toymodel}, based on our hydrodynamics simulations (see Section \ref{subsec:modeling}), and the observed spectral energy distribution from FU Ori, Z CMa, V1735 Cyg and V1057 Cyg (Gramajo et al. 2014). 
The 1.2 millimeter data points, including an upper limit for V1735 Cyg, were observed with the Submillimeter Array, and will be described in a separated paper (P.I.: Dunham M. M.).
}} 
\vspace{0.8cm}
\label{fig:sed}
\end{figure}

Despite the fact that coronagraphic PI observations can provide high angular resolution images with minimized confusion of protostellar emission, it is important to note that the PI images trace surface morphology of disk, rather than column density of gas. 
In the framework of  hydrodynamics simulations, only the latter is typically elucidated.
For example, our simulated  gas column density distributions are presented in Figure \ref{fig:hydrodynamics}.
Radiative transfer models of near infrared reflected light images, based on the simulated density distributions, can provide the link between observations and simulations.
Figure \ref{fig:toymodel} shows a model of our simulated PI images based on the numerical hydrodynamics simulations outlined in Section \ref{subsec:modeling}. 
Our procedure to estimate local scale height self-consistently considered thermal balance, the gravity of the central star, and the self-gravity of the disk. 
The model PI images exhibit several 10$^{2}$ AU spiral arms, or arcs, which have linearly polarized H-band (1.6 $\mu$m) intensity contrasts comparable to our Subaru-HiCIAO observations. 
In addition, our radiative transfer models reproduce the shape of the spectral energy distributions, which are qualitatively similar to previous observations from infrared to millimeter band (Figure \ref{fig:sed}). 
We note that our modeling does not take the cases of binaries into consideration.
We expect that this mainly affect the spectral shapes in the optical and the near infrared bands, which are dominant by the effective stellar photospheric emission (Figure \ref{fig:sed}).

These results demonstrate that the gravitational instabilities in a forming disk is indeed a plausible mechanism for producing the observed structures in the PI images. 
One intriguing aspect from our simulations is that, unlike magnetorotational instabilities (MRI), which creates inhomogeneities on small spatial scales (Balbus \& Hawley 1991), the gravitational instability scenario discussed here more naturally explains the spiral arms on the observed extended spatial scale, and can eject gas clumps or streams, which carry sufficient kinetic energy to pierce the disk and envelope and then leave the system (Basu \& Vorobyov 2012).
In fact, a scaled-up version of the gravitational instability scenario discussed here may have been observed towards OB cluster-forming regions, on $\sim$1 pc scale, although the short evolutionary timescales of massive stars will not permit the later stabilized phase (Liu et al. 2012, 2015).
We consider other explanations such as non-axisymmetric waves on the disk surface, or tidally induced disk features, to be less probable/general but not strictly forbidden, because they may require a large stationary disk to pre-exist in an earlier evolutionary stage, without developing gravitational instability. 
In addition, they cannot explain the cases which do not have external perturbers.

Motivated by the results of numerical hydrodynamics simulations, we naturally interpret the large scale arc to the southeast of V1735 Cyg as an expanding relic ejected from the inner disk due to multiple gravitational interactions (Basu \& Vorobyov 2012). 
The same interpretation may be applied to the ~1000 AU scale elongated structures associated with FU Ori and Z CMA (see Figure \ref{fig:smoothed} for a spatially smoothed image of FU Ori; also see de Leon et al. 2015 for another example). 
In the case of Z CMa, the ejected relics may be further swept up by protostellar wind or jet, and therefore become a narrow feature closely following the edge of the wind/jet (Millan-Gabet \& Monnier 2002; Canovas et al. 2015).

\section{Conclusion and Remarks} \label{sec:summary}
As a summary, the presented Subaru Hi-CIAO observations have demonstrated features in common among four FU Orionis objects in the resolved images: the large-scale asymmetrical structures. 
These structures appear consistent with those produced by our hydrodynamics simulations (Figure \ref{fig:hydrodynamics}), which suggest a strongly unstable phase in the early evolution of protoplanetary disks.
At this unstable stage, the developing gravitational instability in the accreting disk naturally breaks spatial symmetry, creating spiral arms and clumps, which further lead to time variable protostellar accretion and FU Orionis accretion outbursts. 
These dramatic asymmetric features persist through at least several hundred thousand years of the early disk evolution.
The proposed scenario is consistent with the results from a Spitzer and Infrared Space Observatory (ISO) survey of the 10 $\mu$m silicate emission/absorption feature, which proposed that the FU Orionis phase is the link between objects which remain embedded in the circumstellar envelopes (i.e. Class I), and naked (i.e. Class II) disks (Quanz et al. 2007). 
The previous analyses based on far infrared and (sub)millimeter measurements with relatively poor spatial resolutions also claimed the association with circumstellar envelopes (Gramajo et al. 2014). 
Our Subaru-HiCIAO near infrared images are probing circumstellar structures with a more than 100 times improved angular resolution over those previous far infrared and (sub)millimeter observations used for the SED analysis. 
The small innermost working angular scale (0$''$.3) of HiCIAO permitted high angular resolution and high dynamic range images connecting the spatial scales from the circumstellar disk to the envelope. 
These spatially resolved images therefore are presenting a much clearer and robust picture of circumstellar disk and envelope systems than before.
The previous optical and near infrared imaging observations (see Grady et al. 2015, Quanz 2015 for the up to date reviews) did not provide a sample of sources which are undergoing accretion outbursts. 
Our reported four FU Orionis objects thus have provided a very different point of view in the context of star formation. 
In particular, our resolved structures are several times bigger than the spiral arms presented in the previously observed sources like MWC 758 and SAO 206462 (Muto et al. 2012; Grady et al. 2013).
The later are typically on the $<$$\sim$100 AU scales, which are distinct from the large scale arms as presented in our models (Figure \ref{fig:hydrodynamics}, \ref{fig:toymodel}). 
For instance, the disk masses of MWC758 and SAO 206462 are on the order of ~1\% of the stellar mass (Andrews at al. 2011). 
Due to the larger velocity shear on the small scales, it is not easy to trigger gravitational instability for these $<$$\sim$100 AU low mass disks. 
Therefore, the discovered features in common from the presented FU Orionis objects, should be considered a missing piece of an overall picture, which does not only help to understand the FU Orionis objects by themselves, but also the evolutionary track of young stellar objects in general (Baraffe et al. 2012; Dunham \& Vorobyov 2012).
The synergy between the presented observations, and the hydrodynamics simulations with radiative transfer modeling, indicates that the formation and evolution of some, if not all protoplanetary systems can be more dynamic and chaotic then was previously thought.

Finally, our assertion based on numerical simulations, is that the large scale gravitational instability will naturally occur when the disk is fed by a collapsing envelope, provided that the parental cloud has sufficient mass and angular momentum (Vorobyov \& Basu 2010). 
This gravitational instability and associated disk fragmentation will then episodically trigger protostellar accretion outbursts (Figure \ref{fig:hydrodynamics}). 
It is therefore important that all four observed objects show features typical of gravitationally unstable disks. 
Gravitational instability may linger through the quiescent stage, though in the immediate pre-burst and actual burst phases it is expected to be the strongest. 
We note that sources undergoing large-scale gravitational instability but not temporally undergoing protostellar accretion outburst, or even being underluminous, are also reasonable in our proposed scenario. 
However, resolving asymmetric disk structures for nearby (e.g. d$<$450 pc) YSOs requires $<$$\sim$0$''$.1 angular resolution. 
Spiral features of Herbig Ae/Be disks can be resolved thanks to the relatively luminous host stars. 
Otherwise, it can be observed with the sources which are undergoing brightness outburst. 
The optical and near infrared imaging using the 30 m class telescopes in the near future (e.g. TMT, E-ELT), or the ALMA long baseline observations, are required to detect such structures from the lower mass and the more quiescent, embedded young stellar objects.

\vspace{-0.3cm}
\acknowledgements
The observational data were taken using the Subaru telescope. 
We thank the supports of Subaru staffs during our observation runs. 
HBL acknowledge the supports of ASIAA. 
We were supported by the Ministry of Science and Technology (MoST) of Taiwan (Grant Nos. 103-2112- M-001-029 and 104-2119-M-001-018). 
EIV acknowledges the support from the Russian Ministry of Education and Science Grant 3.961.2014/K and RFBR grant 14-02-00719.

{\it Facilities:} \facility{Subaru-HiCIAO}


\appendix 

\section{A. A comparison of derived HiCIAO images without, and without performing polarized stellar halo-subtraction}\label{section:appendix}
\begin{figure}[h]
\hspace{-0.25cm}
\begin{tabular}{ p{10cm}  }
\includegraphics[width=8.7cm]{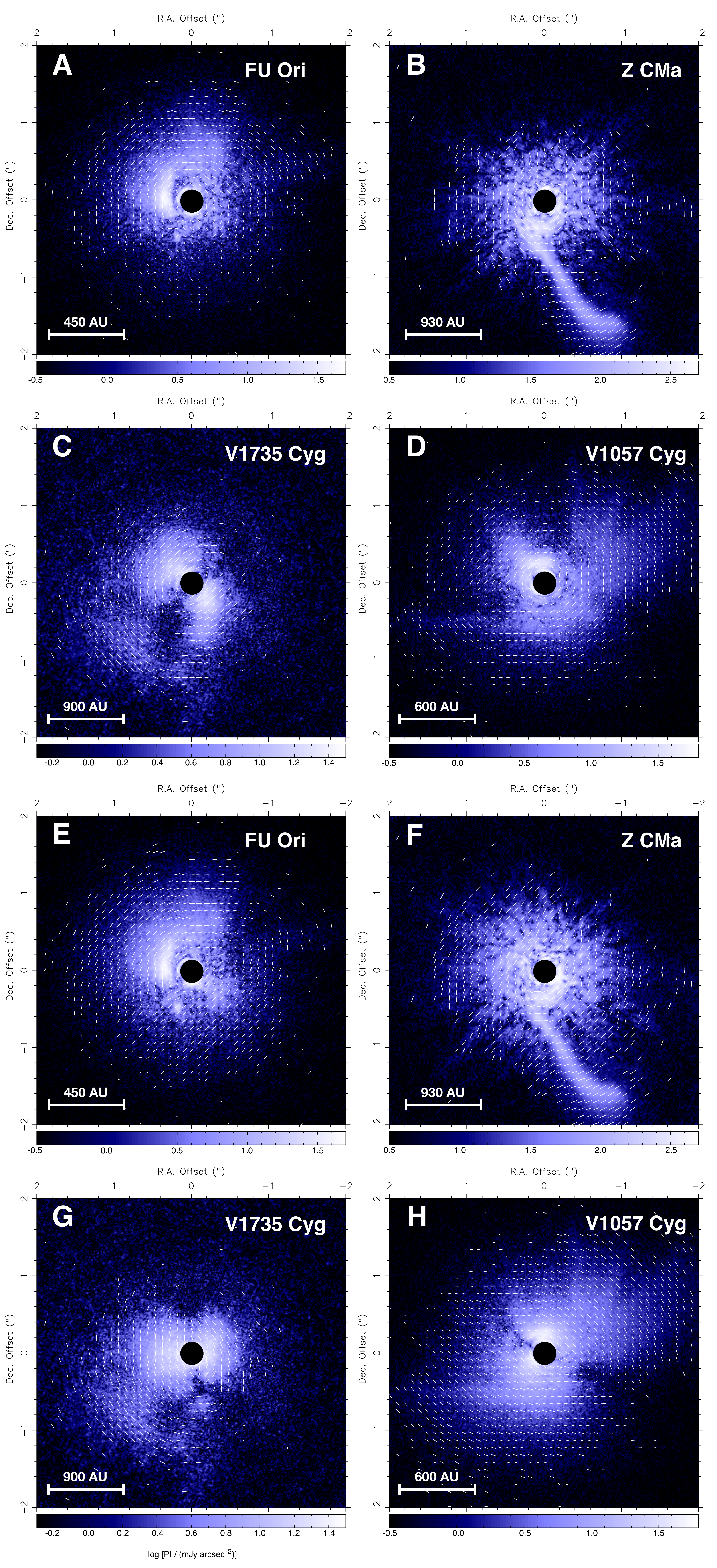}  \\
\end{tabular}
\vspace{-0.0cm}
\caption{\footnotesize{
Polarization images with (A, B, C, D) and without (E, F, G, H) polarized stellar halo-subtraction. Brightness is log-scaled.
}} 
\vspace{0.8cm}
\label{fig:combine}
\end{figure}

We compare the polarization intensity images before and after the polarized stellar halo modeling/subtraction. 
In Figure \ref{fig:combine}, the polarization vectors are plotted in constant length, because it remains difficult to measure the total intensity of the reflected light. 
Before polarized stellar halo modeling/subtraction, polarization vectors appear to have preferred directions on large scales. 
After subtracting models of the polarized stellar halo, polarization vectors tend to align perpendicular to the radial direction.

\vspace{1.5cm}
\begin{thebibliography}{}

\bibitem[Akiyama et al.(2015)]{2015ApJ...802L..17A} Akiyama, E., Muto, T., Kusakabe, N., et al.\ 2015, \apjl, 802, L17

\bibitem[Andrews et al.(2011)]{2011ApJ...732...42A} Andrews, S.~M., Wilner, D.~J., Espaillat, C., et al.\ 2011, \apj, 732, 42

\bibitem[Audard et al.(2014)]{2014prpl.conf..387A} Audard, M., {\'A}brah{\'a}m, P., Dunham, M.~M., et al.\ 2014, Protostars and Planets VI, 387 

\bibitem[Bae et al.(2014)]{2014ApJ...795...61B} Bae, J., Hartmann, L., Zhu, Z., \& Nelson, R.~P.\ 2014, \apj, 795, 61 

\bibitem[Balbus \& Hawley(1991)]{1991ApJ...376..214B} Balbus, S.~A., \& Hawley, J.~F.\ 1991, \apj, 376, 214 

\bibitem[Baraffe et al.(2009)]{2009ApJ...702L..27B} Baraffe, I., Chabrier, G., \& Gallardo, J.\ 2009, \apjl, 702, L27 

\bibitem[Baraffe et al.(2012)]{2012ApJ...756..118B} Baraffe, I., Vorobyov, E., \& Chabrier, G.\ 2012, \apj, 756, 118

\bibitem[Basu \& Vorobyov(2012)]{2012ApJ...750...30B} Basu, S., \& Vorobyov, E.~I.\ 2012, \apj, 750, 30 

\bibitem[Bell \& Lin(1994)]{1994ApJ...427..987B} Bell, K.~R., \& Lin, D.~N.~C.\ 1994, \apj, 427, 987

\bibitem[Bonnell \& Bastien(1992)]{1992ApJ...401L..31B} Bonnell, I., \& Bastien, P.\ 1992, \apjl, 401, L31 

\bibitem[Canovas et al.(2015)]{2015A&A...578L...1C} Canovas, H., Perez, S., Dougados, C., et al.\ 2015, \aap, 578, L1 

\bibitem[Caratti o Garatti et al.(2011)]{2011A&A...526L...1C} Caratti o Garatti, A., Garcia Lopez, R., Scholz, A., et al.\ 2011, \aap, 526, L1

\bibitem[Chabrier \& Baraffe(1997)]{1997A&A...327.1039C} Chabrier, G., \& Baraffe, I.\ 1997, \aap, 327, 1039 

\bibitem[Colella \& Woodward(1984)]{1984JCoPh..54..174C} Colella, P., \& Woodward, P.~R.\ 1984, Journal of Computational Physics, 54, 174 

\bibitem[Dong et al.(2012)]{2012ApJ...760..111D} Dong, R., Hashimoto, J., Rafikov, R., et al.\ 2012, \apj, 760, 111 

\bibitem[Dong et al.(2015)]{2015ApJ...809...93D} Dong, R., Zhu, Z., \& Whitney, B.\ 2015, \apj, 809, 93 

\bibitem[Dunham \& Vorobyov(2012)]{2012ApJ...747...52D} Dunham, M.~M., \& Vorobyov, E.~I.\ 2012, \apj, 747, 52 

\bibitem[de Leon et al.(2015)]{2015ApJ...806L..10D} de Leon, J., Takami, M., Karr, J.~L., et al.\ 2015, \apjl, 806, L10 

\bibitem[Evans et al.(2009)]{2009ApJS..181..321E} Evans, N.~J., II, Dunham, M.~M., J{\o}rgensen, J.~K., et al.\ 2009, \apjs, 181, 321

\bibitem[Follette et al.(2013)]{2013ApJ...767...10F} Follette, K.~B., Tamura, M., Hashimoto, J., et al.\ 2013, \apj, 767, 10 

\bibitem[Galv{\'a}n-Madrid et al.(2015)]{2015ApJ...806L..32G} Galv{\'a}n-Madrid, R., Rodr{\'{\i}}guez, L.~F., Liu, H.~B., et al.\ 2015, \apjl, 806, L32

\bibitem[Goodrich(1987)]{1987PASP...99..116G} Goodrich, R.~W.\ 1987, \pasp, 99, 116 

\bibitem[Grady et al.(2013)]{2013ApJ...762...48G} Grady, C.~A., Muto, T., Hashimoto, J., et al.\ 2013, \apj, 762, 48 

\bibitem[Grady et al.(2015)]{2015Ap&SS.355..253G} Grady, C., Fukagawa, M., Maruta, Y., et al.\ 2015, \apss, 355, 253

\bibitem[Gramajo et al.(2014)]{2014AJ....147..140G} Gramajo, L.~V., Rod{\'o}n, J.~A., \& G{\'o}mez, M.\ 2014, \aj, 147, 140 

\bibitem[Hashimoto et al.(2011)]{2011ApJ...729L..17H} Hashimoto, J., Tamura, M., Muto, T., et al.\ 2011, \apjl, 729, L17 

\bibitem[Hashimoto et al.(2012)]{2012ApJ...758L..19H} Hashimoto, J., Dong, R., Kudo, T., et al.\ 2012, \apjl, 758, L19 

\bibitem[Hartmann \& Kenyon(1996)]{1996ARA&A..34..207H} Hartmann, L., \& Kenyon, S.~J.\ 1996, \araa, 34, 207

\bibitem[Hayano et al.(2004)]{2004SPIE.5490.1088H} Hayano, Y., Saito, Y., Saito, N., et al.\ 2004, \procspie, 5490, 1088 

\bibitem[Herbig(1989)]{1989ESOC...33..233H} Herbig, G.~H.\ 1989, European Southern Observatory Conference and Workshop Proceedings, 33, 233 

\bibitem[Kenyon et al.(1993)]{1993ApJ...414..676K} Kenyon, S.~J., Calvet, N., \& Hartmann, L.\ 1993a, \apj, 414, 676

\bibitem[Kenyon et al.(1993)]{1993ApJ...414..773K} Kenyon, S.~J., Whitney, B.~A., Gomez, M., \& Hartmann, L.\ 1993b, \apj, 414, 773

\bibitem[Kim et al.(1994)]{1994ApJ...422..164K} Kim, S.-H., Martin, P.~G., \& Hendry, P.~D.\ 1994, \apj, 422, 164

\bibitem[Koresko et al.(1991)]{1991AJ....102.2073K} Koresko, C.~D., Beckwith, S.~V.~W., Ghez, A.~M., Matthews, K., \& Neugebauer, G.\ 1991, \aj, 102, 2073 

\bibitem[Kudo et al.(2008)]{2008ApJ...673L..67K} Kudo, T., Tamura, M., Kitamura, Y., et al.\ 2008, \apjl, 673, L67

\bibitem[Kuzuhara et al.(2013)]{2013ApJ...774...11K} Kuzuhara, M., Tamura, M., Kudo, T., et al.\ 2013, \apj, 774, 11 

\bibitem[Lin et al.(1985)]{1985MNRAS.212..105L} Lin, D.~N.~C., Faulkner, J., \& Papaloizou, J.\ 1985, \mnras, 212, 105

\bibitem[Liu et al.(2012)]{2012ApJ...756...10L} Liu, H.~B., Jim{\'e}nez-Serra, I., Ho, P.~T.~P., et al.\ 2012, \apj, 756, 10

\bibitem[Liu et al.(2015)]{2015ApJ...804...37L} Liu, H.~B., Galv{\'a}n-Madrid, R., Jim{\'e}nez-Serra, I., et al.\ 2015, \apj, 804, 37 

\bibitem[Lucy(1999)]{1999A&A...344..282L} Lucy, L.~B.\ 1999, \aap, 344, 282 

\bibitem[Machida et al.(2011)]{2011ApJ...729...42M} Machida, M.~N., Inutsuka, S.-i., \& Matsumoto, T.\ 2011, \apj, 729, 42 

\bibitem[Millan-Gabet \& Monnier(2002)]{2002ApJ...580L.167M} Millan-Gabet, R., \& Monnier, J.~D.\ 2002, \apjl, 580, L167 

\bibitem[Muto et al.(2012)]{2012ApJ...748L..22M} Muto, T., Grady, C.~A., Hashimoto, J., et al.\ 2012, \apjl, 748, L22

\bibitem[Nayakshin(2010)]{2010MNRAS.408L..36N} Nayakshin, S.\ 2010, \mnras, 408, L36 

\bibitem[Nayakshin \& Lodato(2012)]{2012MNRAS.426...70N} Nayakshin, S., \& Lodato, G.\ 2012, \mnras, 426, 70

\bibitem[Pfalzner(2008)]{2008A&A...492..735P} Pfalzner, S.\ 2008, \aap, 492, 735 

\bibitem[Quanz et al.(2007)]{2007ApJ...668..359Q} Quanz, S.~P., Henning, T., Bouwman, J., et al.\ 2007, \apj, 668, 359 

\bibitem[Quanz(2015)]{2015Ap&SS.357..148Q} Quanz, S.~P.\ 2015, \apss, 357, 148

\bibitem[Safron et al.(2015)]{2015ApJ...800L...5S} Safron, E.~J., Fischer, W.~J., Megeath, S.~T., et al.\ 2015, \apjl, 800, L5 

\bibitem[Shu(1977)]{1977ApJ...214..488S} Shu, F.~H.\ 1977, \apj, 214, 488 

\bibitem[Stone \& Norman(1992)]{1992ApJS...80..791S} Stone, J.~M., \& Norman, M.~L.\ 1992, \apjs, 80, 791 

\bibitem[Takami et al.(2013)]{2013ApJ...772..145T} Takami, M., Karr, J.~L., Hashimoto, J., et al.\ 2013, \apj, 772, 145 

\bibitem[Takami et al.(2014)]{2014ApJ...795...71T} Takami, M., Hasegawa, Y., Muto, T., et al.\ 2014, \apj, 795, 71 

\bibitem[Tamura et al.(2006)]{2006SPIE.6269E..0VT} Tamura, M., Hodapp, K., Takami, H., et al.\ 2006, \procspie, 6269, 62690V 

\bibitem[Truelove et al.(1998)]{1998ApJ...495..821T} Truelove, J.~K., Klein, R.~I., McKee, C.~F., et al.\ 1998, \apj, 495, 821 

\bibitem[Vorobyov \& Basu(2007)]{2007MNRAS.381.1009V} Vorobyov, E.~I., \& Basu, S.\ 2007, \mnras, 381, 1009 

\bibitem[Vorobyov \& Basu(2009)]{2009MNRAS.393..822V} Vorobyov, E.~I., \& Basu, S.\ 2009, \mnras, 393, 822 

\bibitem[Vorobyov(2010)]{2010ApJ...713.1059V} Vorobyov, E.~I.\ 2010, \apj, 713, 1059 

\bibitem[Vorobyov \& Basu(2010)]{2010ApJ...719.1896V} Vorobyov, E.~I., \& Basu, S.\ 2010, \apj, 719, 1896 

\bibitem[Vorobyov(2011)]{2011ApJ...728L..45V} Vorobyov, E.~I.\ 2011a, \apjl, 728, L45 

\bibitem[Vorobyov(2011)]{2011ApJ...729..146V} Vorobyov, E.~I.\ 2011b, \apj, 729, 146 

\bibitem[Vorobyov(2013)]{2013A&A...552A.129V} Vorobyov, E.~I.\ 2013, \aap, 552, A129 

\bibitem[Vorobyov et al.(2013)]{2013A&A...557A..35V} Vorobyov, E.~I., Baraffe, I., Harries, T., \& Chabrier, G.\ 2013, \aap, 557, A35 

\bibitem[Vorobyov \& Basu(2015)]{2015ApJ...805..115V} Vorobyov, E.~I., \& Basu, S.\ 2015, \apj, 805, 115

\bibitem[Wang et al.(2004)]{2004ApJ...601L..83W} Wang, H., Apai, D., Henning, T., \& Pascucci, I.\ 2004, \apjl, 601, L83 

\bibitem[White(1979)]{1979ApJ...229..954W} White, R.~L.\ 1979, \apj, 229, 954
 
\bibitem[Whitney et al.(2013)]{2013ApJS..207...30W} Whitney, B.~A., Robitaille, T.~P., Bjorkman, J.~E., et al.\ 2013, \apjs, 207, 30 

\bibitem[Young \& Evans(2005)]{2005ApJ...627..293Y} Young, C.~H., \& Evans, N.~J., II 2005, \apj, 627, 293 

\bibitem[Zhu et al.(2007)]{2007ApJ...669..483Z} Zhu, Z., Hartmann, L., Calvet, N., et al.\ 2007, \apj, 669, 483 

\bibitem[Zhu et al.(2008)]{2008ApJ...684.1281Z} Zhu, Z., Hartmann, L., Calvet, N., et al.\ 2008, \apj, 684, 1281 

\bibitem[Zhu et al.(2009)]{2009ApJ...694L..64Z} Zhu, Z., Espaillat, C., Hinkle, K., et al.\ 2009a, \apjl, 694, L64

\bibitem[Zhu et al.(2009)]{2009ApJ...694.1045Z} Zhu, Z., Hartmann, L., \& Gammie, C.\ 2009b, \apj, 694, 1045

\end{thebibliography}
\end{document}